\documentclass[12pt]{article}

\usepackage[utf8]{inputenc}
\usepackage{amssymb, amsmath, latexsym, amsfonts, amsthm, mathrsfs} 
\usepackage{graphicx}
\usepackage[sorting=none, maxbibnames=10, maxcitenames=10]{biblatex}
\usepackage{verbatim} 
\usepackage[text={6in,9in},centering]{geometry} 
\usepackage{hyperref} 
\hypersetup{
    colorlinks=true,
    linkcolor=red,
    citecolor=red,
    filecolor=magenta,      
    urlcolor=blue,
}
\usepackage[doublespacing]{setspace} 

\usepackage{soul}  
\usepackage{booktabs}   
\usepackage[affil-it]{authblk}

\title{The Enhanced Parameter Estimation (EPE) - A New Calibration Methodology for Building Energy Simulations}
\author[1]{Kris Subbarao \thanks{Corresponding author: KSubbarao@Golden-Analytics.com}}
\author[2]{Srijan Didwania}
\author[3]{T. Agami Reddy}
\author[2]{Marlin Addison}	

\affil[1]{Golden Analytics, Falls Church, Virginia, USA}
\affil[2]{M. S. Addison and Associates, Tempe, Arizona, USA}
\affil[3]{Arizona State University, Tempe, Arizona, USA}

\begin{document}

\date{January 12, 2021}
\maketitle

\begin{abstract}
Buildings rarely perform as designed/simulated and and there are numerous tangible benefits if this gap is reconciled. A new scientific yet pragmatic methodology - called Enhanced Parameter Estimation (EPE) - is proposed that allows physically relevant parameter estimation rather than a blind force-fit to energy use data. It calibrates a rapidly and inexpensively created simulation model of the building in two stages: (a) building shell calibration with the HVAC system is replaced by an by an ideal system that meets the loads. EPE identifies a small number of high-level heat flows in the energy balance, calculates them with specifically tailored individual driving functions, introduces physically significant parameters to best accomplish energy balance, and, estimates the parameters and their uncertainty bounds. Calibration is thus done with corrective heat flows without any arbitrary tuning of input parameters (b) HVAC system calibration with the building shell replaced by a box with only process loads; as many parameters as the data allows are estimated. Calibration accuracy is enhanced by machine learning of the residual errors. The EPE methodology is demonstrated through: a synthetic building and one an actual 75,000 Sq.Ft. building in Pennsylvania. A subsequent paper will provide further details and applications.
\end{abstract}

\section{Introduction}\label{intro} 
Buildings in the United States consume about 40\% of total primary energy resulting in a fifth of total greenhouse gas emissions. Energy efficiency in buildings is a climate change imperative and a significant business opportunity. Electric utilities benefit significantly from energy efficiency through deferral of transmission and distribution system infrastructure upgrades. With increasing penetration of renewables, the conventional power generation profile has changed drastically; it now has to deal with increased variability. Buildings can provide flexibility from the demand side to adapt to variability of generation. To take full advantage of greenhouse gas emission reductions as well as grid flexibility arising from building operation and energy conservation measures, one needs to perform simulations of numerous scenarios. Although, for some limited purposes, such as monitoring and verification (M\&V), fault detection and simple modifications to building operation, one can use empirical/statistical models, for more comprehensive cases properly reconciled physics-based detailed simulations are essential.

Buildings rarely perform as simulated, even after accounting for design vs as-built differences. Hence the need for reconciliation using monitored data. As noted by Reddy \cite{reddy2006}, “Historically, the calibration process has been an art form that inevitably relies on user knowledge, past experience, statistical expertise, engineering judgment, and an abundance of trial and error”. This approach hardly informs design or simulation. The building performance gap – the serious discrepancy between modeled and actual energy performance – has often caused frustration \cite{perfgap} \cite{teleg}. In the context of building energy, the term calibration is pervasively used to denote any method of reconciling simulations with performance. In this paper we identify and distinguish between tuning, calibration and parameter estimation. Most methods fall within the tuning category. 

ASHRAE initiated a research project (RP1051) in 2004 intended to identify the best tools, techniques, approaches, procedures from the existing body of research and to develop a coherent and systematic calibration methodology involving a Monte Carlo approach that includes both “parameter estimation” and determination of the uncertainty in the calibrated simulation. The research was restricted to reconciling detailed building energy simulation programs against utility bills \cite{reddyetal}. Subsequently, several researchers have made computational advances, still relying on a variant of Monte Carlo, resulting in the development of AutoBEM (Automatic Building Energy Model) \cite{autotune} to automate the calibration process. The Autotune methodology uses multiparameter optimization techniques, in combination with data mining-informed artificial intelligence agents, to automatically modify software inputs so that simulation output matches measured data. The approach involves identifying about 150 of the most important parameters and using machine learning algorithms to “learn” successful versus unsuccessful paths to optimization. However, the approach relies on brute-force with little attempt made to analyze the data in a manner that attaches error estimates and informs building design. Further, the problem is intrinsically over-parametrized and this adds a large degree of indeterminacy at a fundamental level. There is little expectation that the reconciled model has captured the right numerical values of the various input parameters.

A solution, in principle, to this problem of a “scientific” gap between simulations and performance data is to perform a true parameter estimation using standard techniques, such as the Levenberg-Marquardt method; see, for example, Chapter 15 of the book “Modeling of Data” \cite{numrec}. This approach also gives an estimate of the errors in the parameters. A brute force approach of estimating the input parameters to the simulation is not feasible, because there are far too many of them. A new reformulation that transforms the problem into one with a manageably few parameters is given in this paper.

Such a view was taken by Sun and Reddy \cite{sun_reddy} who proposed a building calibration methodology, along the lines of parameter estimation, that “involves several distinct concepts, namely, sensitivity analysis (to identify a subset of strong influential variables), identifiability analysis (to determine how many parameters of this subset can be tuned mathematically and identifies which ones are the best candidates), numerical optimization (to determine the numerical values of this best subset of parameters), and uncertainty analysis (to deduce the range of variation of these parameters)”. Some aspects of this study are adopted in the new methodology for HVAC calibration, but there are important differences as described later.

A true parameter estimation method for reconciling an in-house simulation with actual performance, called PSTAR (Primary and Secondary Terms Analysis and Renormalization) was introduced by Subbarao. \cite{pstar}. In conjunction, an experimental procedure called Short-Term Energy Monitoring (STEM) was developed that provides short-term test specifications and analysis to elicit physically relevant parameters \cite{stemres} \cite{stem}. PSTAR provides a method for identifying and computing primary and secondary terms in energy balances, and then introduces certain parameters that are estimated from monitored data to enforce energy balance in actual buildings. PSTAR analysis of heat flows provides a conceptual framework for the method described here.

At this point, it is necessary to have unambiguous terminology. We will assign precise meaning to the terms: tuning, calibration and parameter estimation subsequently. The following Section lays the ground work for that discussion.

\section{Macro Heat Flows: A Simple Case}\label{macrohf}
Working with macro heat flows contributing to energy balance is central to our methodology of reconciling simulations with measured data. A typical simulation, e.g., EnergyPlus, determines micro heat flows reaching, for example, the indoor air node from each of the other nodes (as well as any direct input). It keeps track of a large number of nodes, and thereby a large number of micro heat flows, but does not distinguish between how much of the heat flow is due to the major external drivers such as indoor temperature, outdoor temperature, solar radiation etc. By contrast, we will consider macro heat flows reaching, for example the indoor air node, from all nodes that can be attributed to a specific driving function. Crucially, we develop a method to obtain these heat flows from specialized EnergyPlus simulations.

Let us start with a simple case. Consider a box with walls (floor included) with nominal conductance $U_{wall,nom}$, area $A_{wall}$,  ceiling with nominal conductance $U_{ceiling,nom}$, area $A_{ceiling}$ and windows with nominal conductance $U_{window,nom}$, area $A_{window}$. Assume the heat capacity of all elements to be zero. The box is electrically heated to maintain an inside temperature of $T_{in}(n)$ at time step n, while the ambient temperature is $T_{amb}(n)$. A “simulation” or an expression for the combined energy heat flows gives the heating energy use as 
\begin{multline} \label{simple_ebe}
Q_{sim,nom}(n)= U_{wall,nom} A_{wall} (T_{in}(n)-T_{amb}(n))+ U_{ceiling,nom} A_{ceiling} (T_{in}(n)-T_{amb}(n))+\\
U_{window,nom} A_{window}(T_{in}(n)-T_{amb}(n))
\end{multline}
where the subscript “nom” refers to nominal or first-guess or assumed values.

Compare this with measured heat $Q_{mea}(n)$ (which from physical considerations, under the above-stated assumption of zero heat capacity, will be proportional to $T_{in}(n)-T_{amb}(n))$. It is most likely that the measured heating energy will differ from the simulated heating energy because the assumed input values of $U_{wall,nom}$, $U_{ceiling,nom}$, and $U_{window,nom}$ are estimates and will differ from the actual values. 
We can tune by tweaking individual values to $U_{wall,tune}$, $U_{ceiling,tune}$, and $U_{window,tune}$ such that the resulting $Q_{sim,tune}(n)$ is as close as possible (in the least squares error sense) to $Q_{mea}(n)$ . This can be done in an infinite number of ways, and serves to illustrates the concept of over-parametrization as stated earlier. An attempt to perform estimation of the micro parameters $U_{wall,tune}$, $U_{ceiling,tune}$, and $U_{window,tune}$ from the regression equation
\begin{multline}
Q_{mea}(n) \approx U_{wall,tune} A_{wall}(T_{in}(n)-T_{amb}(n))+ U_{ceiling,tune} A_{ceiling}(T_{in}(n)-T_{amb}(n))+\\U_{window,tune} A_{window}(T_{in}(n)-T_{amb}(n))
\end{multline}
is mathematically impossible.  Only the combined term 
\begin{equation}
    BLC \triangleq U_{wall,tune} A_{wall}+ U_{ceiling,tune} A_{ceiling}+U_{window,tune} A_{window}
    \label{eqn:BLCdef}
\end{equation}
is well-determined. This quantity is called Building Load Coefficient (BLC). In other words, the regression equation should be modified to 
\begin{equation}
Q_{mea}(n) \approx p_{BLC} Q_{sim,nom}(n)
\end{equation}
where $p_{BLC}$ is the single parameter to be estimated. The deviation of the estimate of $p_{BLC}$ from 1 is a measure of the over or under estimation of BLC. This clearly shows the advantage of working with macro heat flows, in this simple case, namely $Q_{sim,nom}(n)$

Consider now the more realistic case when the box elements have varying amount of thermal mass. The energy balance equation is more complex, and will need to consider transient behavior, i.e., past values of temperatures etc. It is no longer clear what specific combination is well-determined. We can surmise that it will involve effects such as admittances rather than U-values. Once again, we seek a formulation that handles these effects implicitly while only exposing a small number of “knobs” to turn such that a true parameter estimation can be done. What these knobs are is considered at length later. Again, we find it is advantageous to work with macro heat flows.

Let us now add an HVAC system to the box. In cases where an overall equipment/system efficiency can realistically capture or account for actual behavior, it is added as an additional knob. For example, the rated COP of a chiller or the full-load efficiency of a boiler. Even for more complex systems, the building envelope parameters estimated above enable us to use the box as a dynamical calorimeter (i.e., an “instrument” that measures the amount of heat introduced or extracted by the HVAC system knowing the driving forces: inside and outside temperatures, solar radiation etc.) to determine heating and cooling provided in each time step. This would then allow us to estimate the HVAC characteristics.

The PSTAR \cite{pstar} approach uses building description to compute zonal admittances and resulting heat flows by an in-house simulation, and then renormalizes (i.e., introduces and estimates parameters such as $p_{BLC}$) the heat flows to achieve the most accurate energy balance. The patent \cite{patent} describes a method to compute heat flows using the simulation program with specialized forcing functions. The method described here represents major improvement over the previous method.

 \section{Tuning, Calibration and Parameter Estimation}\label{sec:whatiscalib}
Some discussion on terminology is warranted considering the fact that calibration is loosely used as an umbrella term for any method to reconcile simulation with performance data. The terms “tuning”, “calibration”, and “parameter estimation” need to be defined more precisely. Tuning is essentially a process involving varying the large number of input parameters either subjectively or based on some expected probability. A set of “best performing inputs” are retained, and an iterative process is followed until the (set of) simulation inputs is deemed satisfactory. Such subjective approaches have intrinsic limitations due to the large number of inputs that can be selected in a numerous different permutations to yield equivalent outputs (as illustrated earlier in this paper). Tuning is currently common in both research and professional studies. 

\begin{figure}[htbp]
\caption{Schematic to illustrate how a calibration function is used to modify the measured output of a simple device e.g., thermometer. The calibration function is determined initially from known outputs through obvious modifications of the flow chart}
\centering
\includegraphics[width=0.8\textwidth]{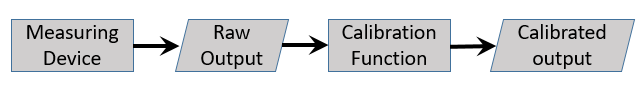}
\label{fig1} 
\end{figure}

\begin{figure}[htbp]
\caption{Schematics of a generalization of the “calibration” process to simulations. Initially, the neural net is trained from measured data through obvious modifications of the flow chart}
\centering
\includegraphics[width=0.8\textwidth]{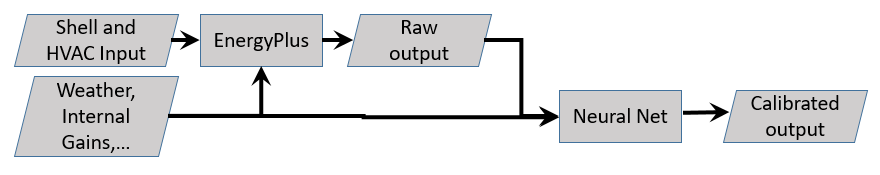}
\label{fig2}
\end{figure}

\begin{figure}[htbp]
\caption{Schematics of how to deploy the Enhanced Parameter Estimation Method. Initially, the parameters are estimated and neural net trained from measured data as described in the text. Note that no modification of shell input parameters of the EnergyPlus model is done. Deviations of simulation inputs from actual are accounted for by corrective heat flows.}
\centering
\includegraphics[width=0.8\textwidth]{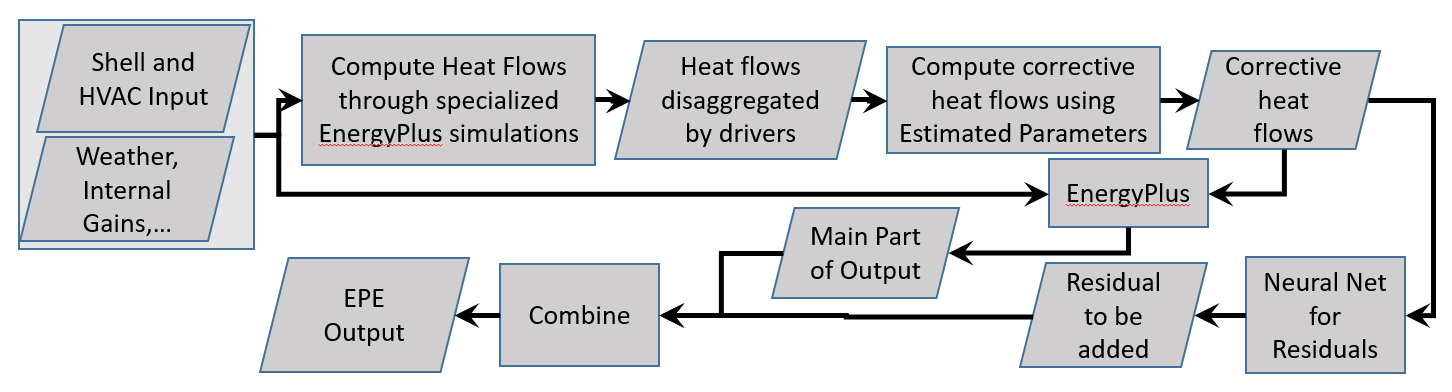}
\label{fig3}
\end{figure}

Calibration is a well-established term in metrology; it essentially means modifying the raw output of a measuring device with a previously-determined correlation to obtain an improved measurement. The metrology-type calibration process is shown in Figure \ref{fig1}. The figure depicts the process that uses the previously determined calibration function say by comparing the instrument raw readings against those from a more accurate reference standard. During training, the known output is used to determine the calibration function. We can think of a simulator such as EnergyPlus as the measuring device, and, its energy use and indoor temperature time series as raw outputs. In this case, there cannot be a simple one-to-one relationship between the raw outputs and calibrated outputs. A generalization is depicted in the Figure \ref{fig2}. EnergyPlus raw output is used as a starting point to empirically obtain improved predictions. For example, using weather and indoor temperature time series as well as energy use time series from EnergyPlus as inputs, we can train a neural net using measured energy use; this can be subsequently used for improved predictions. (This approach must be contrasted with the common method of training a neural net with weather and indoor temperature time series as inputs wherein no simulation is involved). The network parameters have no direct physical interpretation. Further, implementing thermostatic constraints is a problem. We will not pursue this metrology-like calibration approach. What we propose and formulate in detail is an enhanced parameter estimation approach we call EPE (Figure \ref{fig3}). The term "calibration" is so ubiquitous for any reconciliation method, we will use that term loosely.

\section{Enhanced Parameter Estimation (EPE)}\label{sec:epe}
This section describes the mathematical formulation of the method, the multiple specialized EnergyPlus simulations needed, parameters and their estimations, and using machine learning approach to the residuals. Results of applying this methodology to a synthetic building and to a real building are described below.

\subsection{Formulation of the Method}
A simulator such as EnergyPlus performs an energy balance at a number of nodes, in particular at each surface node and each zone air node. Measured performance data on buildings does not permit examining each of these nodes. Let us examine the energy balance at each of the zone air nodes. Consider a linearized version of energy flows in a one zone building; nonlinearities are addressed later through a neural net model for the residuals. The heat flow at time t due to current and past indoor and outdoor temperatures $T_{in}(t')$ and $T_{out}(t')$  can be written as (with the convention that a positive value implies heat gain by the air node) a convolution integral (see \cite{doe2engman}):
\begin{equation}
    \int_{-\infty}^t dt' \, \left[ -V(t-t')T_{in}(t')+W(t-t')T_{out}(t') \right]
    \label{eqn:conv}
\end{equation}
This can be rewritten as a sum of two heat flows $Q_{in}(t)$ and $Q_{BLC}(t)$
\begin{equation}
    Q_{in}(t) =-\int_{-\infty}^t dt' \, V(t-t')T_{in}(t') +LT_{in}(t)
    \label{eqn:QinDef}
\end{equation}
\begin{equation}
    Q_{BLC}(t) = -L \left[ T_{in}(t) - T_{out}^{eff}(t) \right]
    \label{eqn:QBLCgendef}
\end{equation}
\begin{equation}
    T_{out}^{eff}(t) \triangleq \int_{-\infty}^t dt' \, \frac{W(t-t')}{L} T_{out}(t')
    \label{eqn:Touteffdef}
\end{equation}
The quantities $V(t-t')$ and $W(t-t')$ are the appropriate transfer functions. They can be computed from a building description. We do not need to explicitly compute them. If all exterior walls are extremely light, $W(t-t')=L\delta(t-t')$ and the effective outside temperature is equal to the actual instantaneous outside temperature. As the walls become more massive, the effective outside temperature is a weighted average of current and past temperatures, the weights depending on the wall characteristics. For a lightweight wall, the weights decrease rapidly unlike for a heavyweight wall. (It is possible to use only the current outside temperature to define this term but this results in undesirable for heavyweight walls). Sky temperature effects can easily be included by redefining the effective outside temperature:
\begin{equation}
    T_{out}^{eff}(t) \triangleq \int_{-\infty}^t dt' \, \left[ \frac{W(t-t')}{L} T_{out}(t') + \frac{W_{sky}(t-t')}{L} T_{sky}(t') \right]
    \label{eqn:Teffdefinclsky}
\end{equation}

Note that a continuous time formulation is used. It is straightforward to switch between continuous and discrete formulation as convenient. 

We will now add $Q_{sun}(t)$ to account for solar gains, and $Q_{LEP}(t)$ to account for gains from light, equipment and people that appear as heat gain at the indoor-air node. Note in particular that the term $Q_{LEP}(t)$ is the heat gain by the air node after accounting for the delayed release of the gains and any loss through the walls without appearing as gains at the air node. Additional energy flow terms such as infiltration, ventilation and ground flow will be included when necessary. For the time being, for pedagogical simplicity, we have set them to zero.
The heating/cooling load at the zone air node can be written as:
\begin{equation}
    Q_{load}(t) \triangleq Q_{BLC}(t)+Q_{in}(t)+Q_{sun}(t)+Q_{LEP}(t)
    \label{eqn:load}
\end{equation}
Let us denote by $Q_{HC}(t)$  the heating/cooling delivered to the air node by the HVAC system. Finally, the energy balance equation for the air node is:
\begin{equation}
    Q_{load}(t) + Q_{HC}(t) = 0
    \label{eqn:ebe}
\end{equation}
The building load $Q_{HC}(t)$ is met by energy $E(t)$ (electricity and/or gas collectively) that is a function of $Q_{HC}(t)$, weather variables such as outdoor temperature and humidity collectively denoted by Weather(t), and a set of system parameters such as COP, part load efficiency etc. collectively denoted by \{$\alpha$\}:
\begin{equation}
    E(t) = Function(Q_{HC}(t), Weather(t), \{\alpha\})
    \label{eqn:ener}
\end{equation}
The energy use $E(t)$ is an output of the simulation. Our main problem is how best to reconcile simulated values of $E(t)$ with the measured values.

The basic methodology is best explained through a one zone building. Variations can be addressed as needed.
\newline
\newline
We start with:
\begin{itemize}
  \item 
 An inexpensively created EnergyPlus simulation model, called audit model, that is the one to be reconciled with performance data

  \item{The available performance data consists of time series of}
    \begin{itemize}
    \item{Indoor temperature $T_{in,measured}(t)$}
    \item{Weather (outdoor temperature, solar radiation,...) collectively denoted by $Weather(t)$}
    \item{Heat input from lights, equipment and people $LEP(t)$}
    \item{Energy Use (electricity, gas) collectively denoted by $E(t)$}
    \item{Additionally, short-term data for the actual heat input $Q_{HC}^{mea}(t)$ (may be negative) possibly through flow $\Delta$T type measurements}
    \end{itemize}
\end{itemize}
The analysis consists of two stages. In Stage 1, the building shell related parameters are estimated; HVAC system is replaced by an ideal system that simply delivers the required heating/cooling. In Stage 2, the HVAC related parameters are estimated; the building shell is replaced by a box with only process loads.
\newline
\newline
Stage 1: Building Shell parameters 
\begin{itemize}
    \item 
    Perform an EnergyPlus simulation of the audit building, for the period for which $Q_{measured}(t)$ is available using $Weather(t)$, $LEP(t)$ with $T_{in,measured}(t)$ as the set-points. Denote the resulting load by $Q_{1}(t)$. If the audit building shell accurately represents the real building shell (and the simulator is accurate), $Q_{1}(t)$ would be equal to $Q_{measured}(t)$. We do not expect this to be the case and now need to reconcile simulated and actual performances.
    \item
    Perform a second EnergyPlus run, this time with the solar radiation set to 0, and obtain the required input $Q_{2}(t)$. The heat flow
    \begin{equation}  
    Q_{sun}(t) \triangleq  Q_{2}(t) - Q_{1}(t)
    \label{eqn:Qsundef}
    \end{equation}
    is identified as the heat input to the indoor air node due to solar radiation after accounting for any delayed release.
    \item
    Perform a third EnergyPlus run, this time with the indoor temperature fixed at a constant value $T_{in,fixed1}$ and the solar radiation set to 0, to obtain the required input $Q_{3}(t)$.
    \item
    Perform a fourth EnergyPlus run, this time with the indoor temperature fixed at a constant value $T_{in,fixed2}$ and the solar radiation set to 0, to obtain the required input $Q_{4}(t)$.
    \item
    Perform a fifth EnergyPlus run, this time with the indoor temperature fixed at a constant value $T_{in,fixed1}$ and the solar radiation as well as Light, Equipment and People heat input set to 0, to obtain the required input $Q_{5}(t)$.
    \item
    Obtain the following heat flows:
    \begin{equation}
    Q_{in}(t) \triangleq  -Q_{2}(t)+Q_{3}(t) \frac{T_{in,measured}(t)-T_{in,fixed2}}{T_{in,fixed1}-T_{in,fixed2}} +\\Q_{4}(t) \frac{T_{in,fixed1}-T_{in,measured}(t)}{T_{in,fixed1}-T_{in,fixed2}}
    \label{eqn:Qindef}
    \end{equation}
    \begin{equation}
    Q_{BLC}(t) \triangleq  -Q_{2}(t) + Q_{3}(t)-Q_{5}(t) - Q_{in}(t)
    \label{eqn:QBLCdef}
    \end{equation}
    \begin{equation}
    Q_{LEP}(t) \triangleq  -Q_{3}(t) + Q_{5}(t)
    \label{eqn:QLEPdef}
    \end{equation}
    Eqs \ref{eqn:Qsundef} and \ref{eqn:QLEPdef} make intuitive sense. Note that $Q_{LEP}(t)$ is the effect of ${LEP}(t)$ after accounting for any time delays and losses. The term $Q_{in}(t)$ is the heat flow to the air node based on the history of indoor temperatures. $Q_{BLC}(t)$ is the heat flow to the air node based on the current indoor temperature and an effective outdoor temperature that is based on the history of outdoor temperatures (including contributions from the history of sky temperature depression)
    \item Introduce parameters to be estimated and estimate them
    \newline
    By construction, the following equation is satisfied:
    \begin{equation}
    Q_{BLC}(t)+Q_{in}(t)+Q_{sun}(t)+Q_{LEP}(t)+Q_{1}(t)=0
    \label{eqn:QIdentdef}
    \end{equation}
    We now bring in $Q_{HC}^{mea}(t)$, and require that Eq.\ref{eqn:QIdentdef} be satisfied with $Q_{HC}^{mea}(t)$, not $Q_{1}(t)$. We modify the rest of the heat flows by introducing parameters and estimate them to satisfy the equation in the least squares error sense. The simplest modification is:
    \begin{equation}
    p_{BLC}Q_{BLC}(t)+p_{in}Q_{in}(t)+p_{sun}Q_{sun}(t))+p_{LEP}Q_{LEP}(t)+Q_{HC}^{mea}(t)\approx 0
    \label{eqn:Qlinmoddef}
    \end{equation}

This results in a simple linear regression problem. These parameters change only the scale of a heat flow but not the shape. To change the shape we introduce additional parameters. A heat flow, e.g., $Q_{sun}(t)$ is modified through a transfer function as follows:
    \begin{equation}
    Q_{sun}^{mod}=p_{sun}Q_{sun}(t))+Q_{sun}^{TF}(t)
    \label{eqn:Qsunmoddef}
    \end{equation}
    where
    \begin{equation}
    Q_{sun}^{TF}(t)=\alpha_{sun} Q_{sun}^{TF}(t-1)+\beta_{sun} (Q_{sun}(t)-Q_{sun}(t-1))
    \label{eqn:Qsuntfdef}
    \end{equation}
The new energy balance equation is    
\begin{multline}
    p_{BLC}Q_{BLC}(t) + p_{in}Q_{in}(t) + p_{sun}Q_{sun}(t) + p_{LEP}Q_{LEP}(t) + \\Q_{TF,sun}(t) + Q_{HC}^{mea}(t) \approx 0
    \label{eqn:regressimproved}
\end{multline}    
    This introduces 2 additional parameters for the solar heat flow $\alpha_{sun}$, $\beta_{sun}$. Similar parameters can be introduced to the other heat flows. We have to be judicious in introducing additional parameters through an error analysis to ensure that the data supports their estimation. We now have a nonlinear regression problem. This can be solved using Levenberg-Marquardt method.
    Finally, it is obvious that the simulated building would behave close to the real building if we add the following process load to the simulated building:
    \begin{multline}
    Q_{correction}(t) \triangleq (p_{BLC}-1)Q_{BLC}(t)+(p_{in}-1)Q_{in}(t)+(p_{sun}-1)Q_{sun}(t))+\\
    (p_{LEP}-1)Q_{LEP}(t)+Q_{sun}^{TF}(t)
    \label{eqn:QCorrection}
    \end{multline}
    Similar modifications are needed to accommodate transfer functions for any of the heat flows. One can create any number of corrective heat flows (for example, one due to sky temperature depression), but judicious, parsimonious choice based on experimentation is necessary. The EnergyPlus simulation we started with is modified by adding $Q_{correction}(t)$ as a process heat. The resulting output is EPE's calibrated output. EPE accomplishes shell calibration without any modification of the simulation inputs (as noted in the Introduction section, data does not support any systematic modification of the vastly undetermined problem); the deviation of the simulated building from the actual is corrected by introducing corrective heat flows.
\item
Examine the parameters in light of their physical significance
\newline
The deviation of, for example, $p_{BLC}$ from $1$ is a measure of the extent to which the initial estimate of BLC is over or under the actual estimate. The simulation input file can be reexamined in light of such deviations, but there is no systematic way to assign the correction to individual walls, windows etc. (Large deviations may point to possible bugs in inputs; such erroneous inputs should be obviously fixed.) Similarly all the other parameters should be examined. 
 \item
 Develop a neural net for the residuals
 \newline
 With the various heat flows as inputs and residuals as output, a neural net is trained to improve the energy balance
\end{itemize}

Stage 2: HVAC System
\begin{itemize}
    \item 
    Replace the building shell by a process load whose time series is given by:
    \begin{equation}
        Q_{reconciled}(t)\triangleq-Q_{BLC}^{mod}(t)-Q_{in}^{mod}(t)-Q_{sun}^{mod}(t)-Q_{LEP}^{mod}(t)+Q_{residual}(t)
        \label{eqn:Qrecon}
    \end{equation}
    Specialized simulations are done to obtain the heat flows which are then modified to obtain the process load. (we can, of course, use $Q_{measured}(t)$ directly for the periods when it is available)
    \item
    Perform an EnergyPlus simulation with the audit HVAC system and the process load to determine the simulated energy use $E_{sim}(t)$.
    \newline
    If the audit description of the HVAC system is accurate, then $E_{sim}(t)$ will be close to $E_{measured}(t)$. This is generally not expected.
    \item
    Identify the parameters to be estimated and estimate them
    \newline
    In some ways this is similar to the shell problem but in many ways, there are important differences. First of all, each time step is essentially independent of the previous time steps (unless one is considering minute-by-minute control sequence, or storage elements as part of the HVAC system). There are as for the shell a large number of parameters and not all can be determined with whole building data.
\end{itemize}

It is important to note that we use hourly (or subhourly) data for temperatures, weather, and energy. This is a consistent match with a simulator that works with such inputs and outputs. Such data is available from energy management systems, smart meters and if necessary from nearby weather station. There is a disconnect in calibrating hourly simulation with monthly data (highly overparametrized and so a large number of diverse input variable sets will yield very good fits to utility bills). Unfortunately, this is quite common. In \cite{autotune}, the authors calibrate an EnergyPlus simulation model using hourly data, but without indoor zone temperatures. The thermal mass behavior in response to indoor temperature variations is important in terms of the relative magnitude of internal heat flows and neglecting this cannot yield proper calibrated models.
\newline We emphasize that EPE accomplishes shell calibration without any modification of the simulation inputs (as noted in the Introduction section, data does not support any systematic modification of the vastly undetermined problem); the deviation of the simulated building from the actual is corrected by introducing corrective heat flow (Eq. \ref{eqn:QCorrection}). One can come up with ways to modify the EnergyPlus input so that the resulting $p_{BLC}$ and other similar parameters are (close to) 1. For example, we can multiply all exterior wall layer conductivities by $p_{BLC}$. Any such revision can only be justified with additional component-level data.

\subsection {Synthetic Building Example:}
In this section, we will demonstrate the multi-stage methodology for a synthetic building i.e., a software model of a building from which we generate simulated data and use it as the “real” data. The “real” building is then taken to be the one simulated in EnergyPlus to give hourly time series of indoor temperatures and delivered heating and cooling energy in step 1 and electrical consumption by HVAC system in step 2. In this section, the terms “measured” and “real” refer to this building. 

For this study, the "real" building is the 53,600 sq. ft. building with normal construction which corresponds to the medium office building prototype selected by USDOE for work related to ASHARE 90.1 standard development. The audit building has been intentionally modified so as to have different insulation, mass and solar characteristics; more specifically it has more insulation, more mass (added concrete wall layer) and smaller solar heat gain coefficient. Two locations were selected for analyzing such a building: Phoenix, AZ and Philadelphia, PA. A more detailed study was done for the Phoenix location. 

Note that evaluation with a synthetic building allows us to vary the inputs at will and evaluate the accuracy of the Enhanced Parameter Estimation process with more certainty.
Step 1: Obtain the measured (hourly) values of indoor temperature $T_{in}(t)$ and delivered heating/cooling $Q_{HC}^{mea}(t)$ for the "real" building
Step 2: Run the five EnergyPlus simulations of the audit building as specified and determine the four heat flows $Q_{BLC}(t)$,$Q_{in}(t)$,$Q_{sun}(t)$,$Q_{LEP}(t)$. 
\begin{figure}[htbp]
\caption{The heat flow $Q_{mea}(t)$ is measured and the rest computed (synthetic building). }
\centering
\includegraphics[width=0.8\textwidth]{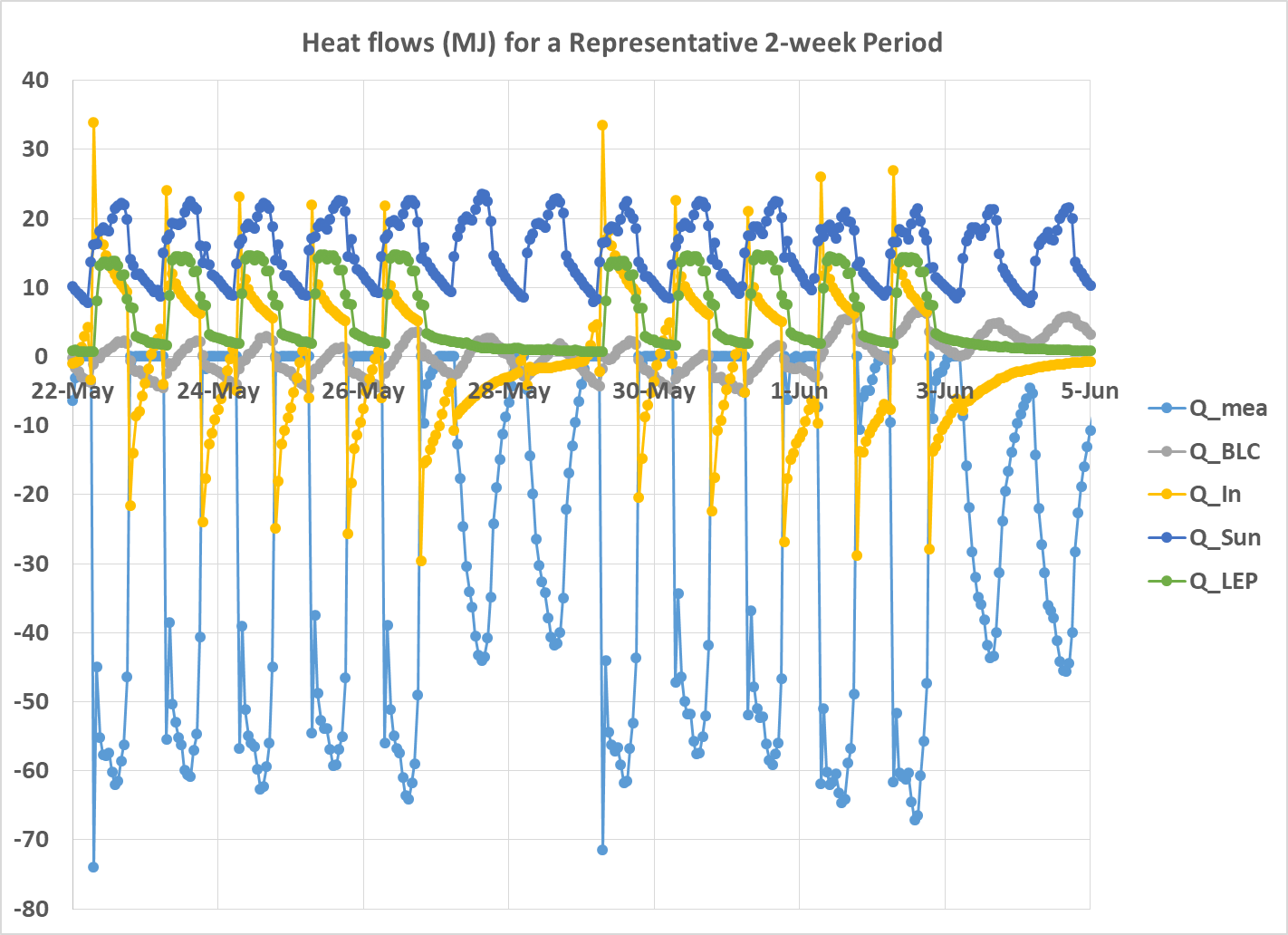}
\label{fig4}
\end{figure} 
These and the “measured” heating/cooling energy delivered are shown in Figure \ref{fig4}. The period of analysis covered two months: May-June. For clarity, only a representative 2-week period is shown in the figure. If the audit building were identical to the real building (and if EnergyPlus is a fully accurate simulator) the five heat flows would add up to 0, or equivalently the negative of the sum of the four computed flows would equal the delivered heating/cooling energy amounts. 
\begin{figure}[htbp]
\caption{The delivered heating/cooling energy flows (equal to the negative sum of the four heat flows) before and after the best fit, compared to measured values}
\centering
\includegraphics[width=0.8\textwidth]{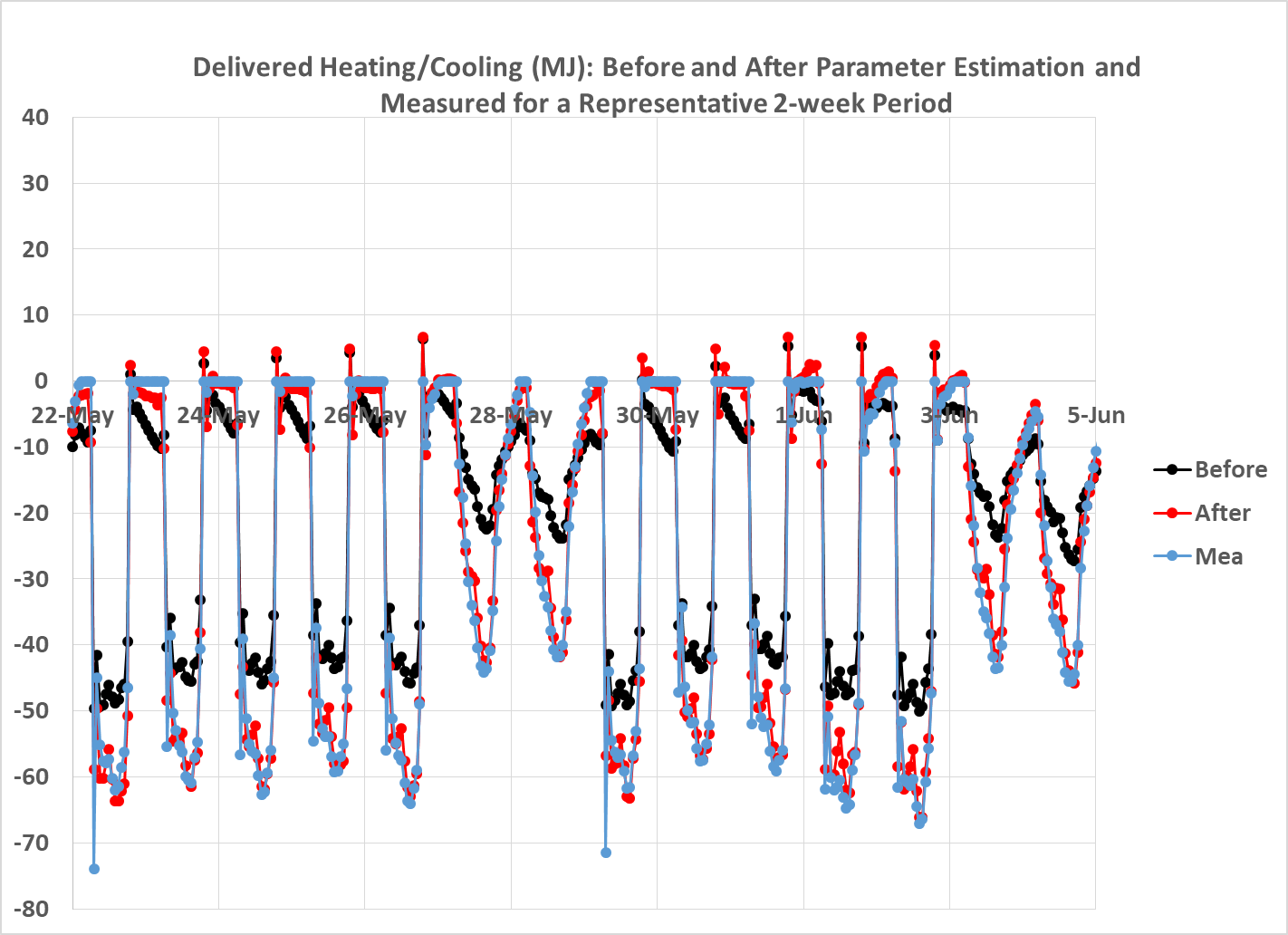}
\label{fig5}
\end{figure} This hourly sequence of negative values of the sum is labeled “Before” in Figure \ref{fig5}. 

Step 3: Perform a non-linear least squares best fit. The negative sum after the best fit is also shown in Figure \ref{fig5}. Note the improvement. The “After” plot is much closer to the “mea” and captures peaks and valleys better. The parameters from the best fit are shown in Table \ref{tab:parphx}. The errors quoted are purely statistical. (To avoid collinearity, $p_{LEP}$ was set to 1 and not estimated in these runs. Such correlations can often be avoided by choosing suitable data windows. This issue needs further study).

\begin{table}[htbp]
  \centering
  \caption{The Estimated parameters (May-June data)}
    \begin{tabular}{|p{4.5em}|p{7.315em}|p{7.315em}|}
    
    \midrule
    $p_{BLC}$   & 1.48$\pm$0.02 \\
    \midrule
    $p_{in}$    & 0.63$\pm$0.05 \\
    \midrule
    $p_{sun}$   & 1.29$\pm$0.01 \\
    \midrule
    $p_{LEP}$   & 1 \\
    \midrule
    $p_{in,phase}$  & 0.43$\pm$0.04 \\
    \midrule
    $p_{TF,in}$  & 0.94$\pm$0.01 \\
    \midrule
    $p_{sun,phase}$  & 1.35$\pm$0.02 \\
    \midrule
    $p_{TF,sun}$  & 0.988$\pm$0.002 \\
    \bottomrule
    \end{tabular}%
  \label{tab:parphx}%
\end{table}%
Step 4: Identify a neural net model for the residuals of the best fit least squares model. To avoid confusion, let us call this PE\_Residuals (for Parameter Estimation Residuals). With the usual caveats, we proceed to improve the estimate by using neural nets. The PE\_residuals show some residual pattern not captured by the parameter estimation process. With the six heat flows $Q_{BLC}(t)$, $Q_{in}(t)$, $Q_{sun}(t)$, $Q_{LEP}(t)$, $Q_{TF,in}(t)$, $Q_{TF,sun}(t)$ as inputs and PE\_Residuals as output with 9 hidden nodes, a multilayer perceptron network was trained on the PE\_Residuals from the May-June data. The fitted PE\_Residuals are now added to the delivered heating/cooling from the parameter estimation step to improve the estimate. The improvement is significant as can be seen from a scatter plot, Figure \ref{fig6}, of the three heat flows vs measured flow:
\begin{figure}[htbp]
\caption{Scatter plot showing the extent to which the Enhanced Parameter Fit process improves the agreement with measured data (May-June data)}
\centering
\includegraphics[width=0.8\textwidth]{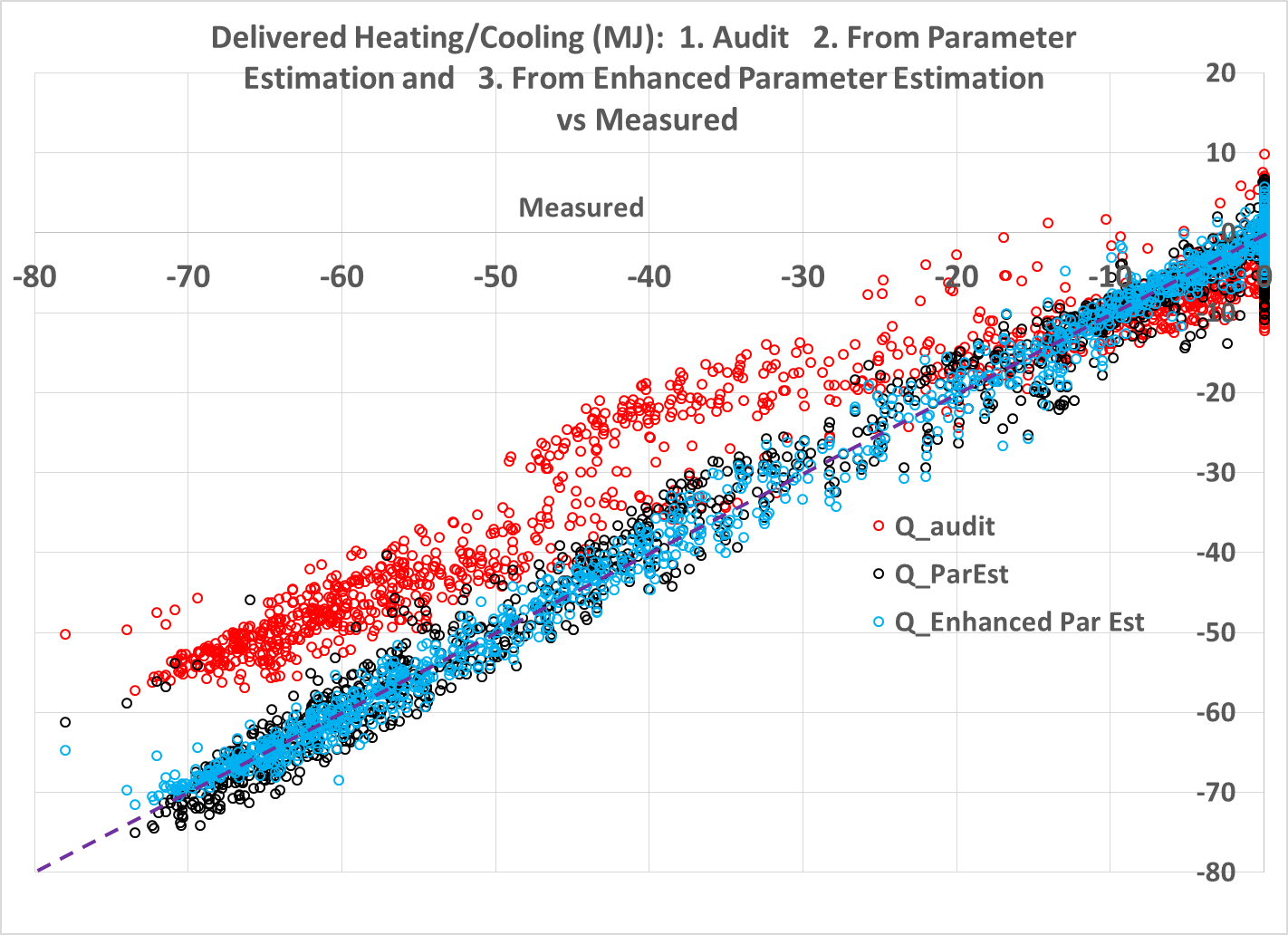}
\label{fig6}
\end{figure} 

\begin{table}[htbp]
  \centering
  \caption{Goodness-of-fit statistics of the EPE process for the synthetic building for the period May-June}
    \begin{tabular}{|p{4.5em}|p{7.315em}|p{7.315em}|p{7.315em}|}
    \toprule
    \multicolumn{1}{|r|}{} & Audit Building & After Parameter Estimation & After Parameter Estimation and Residual Neural Net (Enhanced Parameter Estimation) \\
    \midrule
    Mean Bias Error & 5.13 MJ & -0.02 MJ & -0.05 MJ \\
    \midrule
    Root Mean Square Error & 10.33 MJ & 3.23 MJ & 2.47 MJ \\
    \bottomrule
    \end{tabular}%
  \label{tab:err}%
\end{table}%

The mean bias error and RMS error are shown in Table \ref{tab:err}. Note that after the parameter estimation step, the mean bias error is essentially zero after calibration while the RMSE value has been reduced three-fold. The RMSE is further reduced (from 3.23 MJ to 2.47 MJ) after the neural net fitting procedure is applied.

One of the thorny problems in applying neural networks to time series is handling past values of driving functions. In our case, this is a non-issue since the inputs are heat flows that already incorporate the history of driving functions. Thus, each row is treated independently. The time series problem has been reduced to an independent case problem! 

\subsection{Physical Significance of the Parameters}
The parameters have physical significance which helps us better understand the performance of the building. 
The parameter $p_{BLC}$ was estimated to be 1.48. This implies that the building load coefficient of the actual building is roughly 48\% higher than that of the audit building. We cannot go beyond this to determine which components (walls or windows etc.) are causing this discrepancy; reasonable conjectures would be circumstance specific. Since the “real” building is synthetic, we know the actual numerical values of the parameters . The differences between the actual and the audit building were such that the load coefficient of the actual building is roughly 40\% higher; this is in good agreement with our parameter estimation. The value of the load coefficient is somewhat dependent on the treatment of convection and radiation at the interior and exterior surfaces and this must be resolved if the differences are to be reconciled further. 
 Similarly, $p_{sun}$ = 1.29 implies that the actual building has roughly 29\% higher solar gains. The solar heat gain coefficient of the windows of the actual building was about 47\% higher; we need to consider also opaque gains. Similarly $p_{in}$ = 0.63 means that the actual building has less effective thermal mass; this is consistent with the differences between audit and actual building inputs to EnergyPlus. The other parameters modify the delay characteristics of solar gains and of heat flows due to indoor fluctuations beyond what is in the audit description.
\subsection{HVAC System}
So far, we have used May-June data with hourly delivered heating/cooling data available to us. This enabled us to estimate envelope parameters as well as train a neural net to address the envelope characteristics. We now will use the EPE methodology to predict loads on the mechanical systems, and thereby perform HVAC reconciliation.  
If the delivered heating/cooling data is available, for example, through a BTU meter, we can simply use that data to analyze the HVAC system; in particular, we could estimate certain HVAC parameters. We can visualize a use-case situation where the needed instrumentation is installed temporarily and then removed to reduce operating costs, while being able to keep monitoring the HVAC performance continuously. The proposed procedure involving enhanced parameter estimation would permit using the building envelope as a calorimeter, essentially as a substitute for flow-$\Delta$T type instrumentation. We will discuss this aspect next.
Suppose we have July-August data with indoor temperatures and HVAC electricity use. We can use the indoor temperature data in EnergyPlus to compute the four heat flows as described in the previous section: $Q_{BLC}(t)$, $Q_{in}(t)$, $Q_{sun}(t)$, $Q_{LEP}(t)$ for the period July-August. From the previously estimated $p_{TF,in}$ and $p_{TF,sun}$ parameters, we can determine the two additional flows $Q_{TF,in}(t)$, $Q_{TF,sun}(t)$. We can now apply Eq. \ref{eqn:regressimproved} and include the hourly residual time series data from the neural net to get the best estimate of delivered cooling:
\begin{multline}
        Q_{HVAC}^{estimate}(t)=p_{BLC}Q_{BLC}(t) + p_{in}Q_{in}(t) + p_{sun}Q_{sun}(t) + p_{LEP}Q_{LEP}(t) + \\ p_{TF,sun2}Q_{TF,sun}(p_{TF,sun1},Q_{sun}(t)) + p_{TF,in2}Q_{TF,in}(p_{TF,in1},Q_{in}(t)) +NN_{Res}
        \label{eqn:delivered}
\end{multline}
The term $NN_{Res}$ represents the residual time series estimated from the neural net.

We can now run an EnergyPlus simulation with the estimated loads as process load inputs to an isolated box; the building envelope is replaced by this isolated box for this simulation. Energy Plus takes the load time series and outputs electricity use time series using the user-specified input for the HVAC parameters - COP under rated conditions, and part load, and off-rated performance. We can now perform a non-linear least squares estimation of the HVAC parameters. To do this using software packages requires repeated calls to EnergyPlus. It is important to recognize that the data permits estimating a very small number of HVAC parameters, and these should be chosen judiciously.

\begin{figure}[htbp]
\caption{Search process for the best estimate for cooling system COP that minimizes RMSE}
\centering
\includegraphics[width=0.8\textwidth]{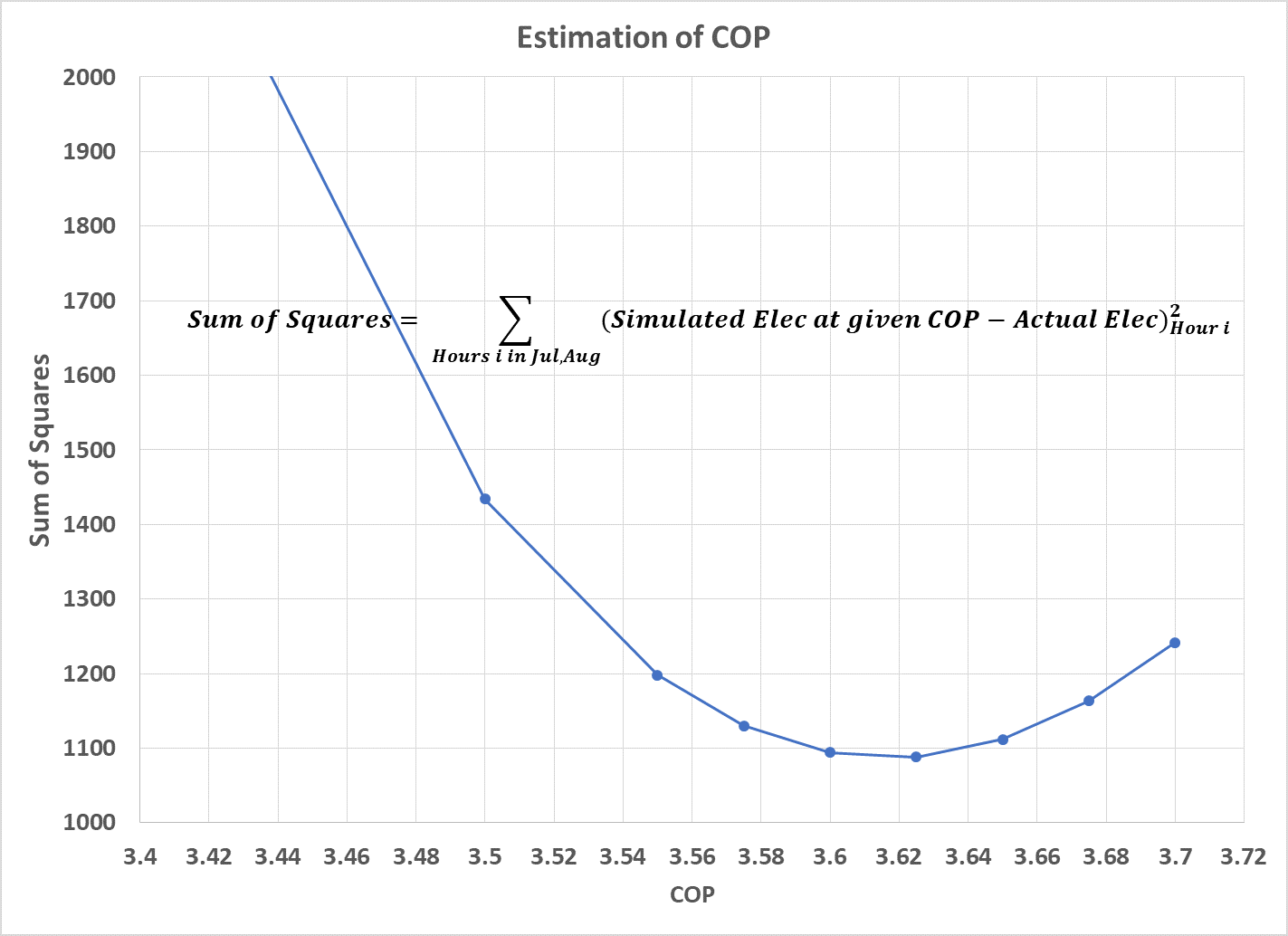}
\label{fig7}
\end{figure}

To illustrate the general approach,  we assume that the only parameter to be estimated is the COP under rated conditions and leave part-load and off-rated corrections at their default values. We can run EnergyPlus for a range of values of COP. We compute the RMS error as a function of rated COP and identify the value at the minimum error; this is a manual process for accomplishing the nonlinear estimation. This process of searching for a COP values which results in minimum RMSE is shown in Figure \ref{fig7}.
\newline
The rated COP that minimizes the error turned out to be 3.625. Because the real building is in fact synthetic, we know the real COP value was 3.5. The difference is due, in addition to uncertainties of the methodology, mainly to an artifact of EnergyPlus. For the building we can deduce the electricity time series from EnergyPlus in two ways: One is via the normal simulation. The second is to first get the cooling loads (with the right indoor temperatures specified) and introduce them to another simulation with the shell replaced by an isolated box. The two are expected to be essentially identical; however, we found a 2\% increase for the isolated box. This suggests that for some unidentified reason (to be investigated), EnergyPlus increases the load by 2\% while simulating the box. To match the normal electricity with a 2\% higher load use requires about a 2\% increase in COP.

\subsection{How repeatable are the parameter estimates}

To evaluate whether the parameter estimates are impacted by climate, we have repeated the analyses for the same months (May-June) in two different climates: Philadelphia and Phoenix. The results are shown in Table \ref{tab:table3}. The agreement is generally satisfactory considering we do not expect them to be identical. As is to be expected from physical considerations, differences in sky temperature, ground reflectivity between the two cities affect the parameters. 
\begin{table}[htbp]
  \centering
  \caption{Estimated Parameters in two different climates along with uncertainty values}
    \begin{tabular}{|p{4.065em}|p{7.315em}|p{7.315em}|}
    \multicolumn{1}{r}{} & \multicolumn{1}{l}{Philadelphia} & \multicolumn{1}{l}{Phoenix} \\
    \midrule
    $p_{BLC}$  & 1.57$\pm$0.03 & 1.48$\pm$0.02 \\
    \midrule
    $p_{in}$   & 0.69$\pm$0.02 & 0.63$\pm$0.05 \\
    \midrule
    $p_{sun}$  & 1.42$\pm$0.01 & 1.29$\pm$0.01 \\
    \midrule
    $p_{LEP}$  & \multicolumn{1}{r|}{1} & \multicolumn{1}{r|}{1} \\
    \midrule
    $p_{in,phase}$ & 0.45$\pm$0.02 & 0.43$\pm$0.04 \\
    \midrule
    $p_{TF,in}$ & 0.70$\pm$0.02 & 0.94$\pm$0.01 \\
    \midrule
    $p_{sun,phase}$ & 1.37$\pm$0.03 & 1.35$\pm$0.02 \\
    \midrule
    $p_{TF,sun}$ & 0.954$\pm$0.002 & 0.988$\pm$0.002 \\
    \bottomrule
    \end{tabular}%
  \label{tab:table3}%
\end{table}%
\subsection{Actual Building In Philadelphia}
For this study, we have identified a building, located in the Philadelphia Navy Yard, for which an EnergyPlus simulation input model was also developed from a previous study carried out by Ke Xu. \cite{KeXu}. Building 101 was initially built in 1911 as a marine barrack with the gross building floor area about 75,000 ft$^2$. This was a brick building with double-pane windows and underwent a major HVAC system renovation in 1999. The building had 3 above grade floors and one Basement floor which primarily served as mechanical and storage rooms. The HVAC system in the building comprised of three variable-air-volume (VAV) air handling units (AHUs) with total design airflow of 53,240 CFM. The terminal units have hydraulic reheat coils to meet the heat / reheat loads which are served by a 1,632 MBH gas-fired boiler. The AHUs are served by DX cooling coils. The boiler supplies 180$^o$F hot water to AHU heating coils, Terminal Box reheat coils, and hot water radiators. The attic spaces are heated by hot water unit heaters. Further details can be found in \cite{KeXu}.
For our study, we had access to the 2016 end-use energy data for each minute, which included lighting electric, AHU fans electric, DX Coils electric, pumps electric, total building electric and miscellaneous equipment electric energy as residual. Gas consumption data by the boilers were also available for each minute. We carried out the analysis at an hourly level. The weather file for this location, for the year 2016, was obtained from Whitebox Technologies \cite{whitebox}.

Real buildings, unlike synthetic buildings, present unique challenges; the EPE methodology should be adapted to meet the challenges. Such inevitable complexities in dealing with real buildings should not detract from the methodology itself.
\begin{figure}[htbp]
\caption{Measured data for the actual building. Only a representative winter period is shown}
\centering
\includegraphics[width=0.8\textwidth]{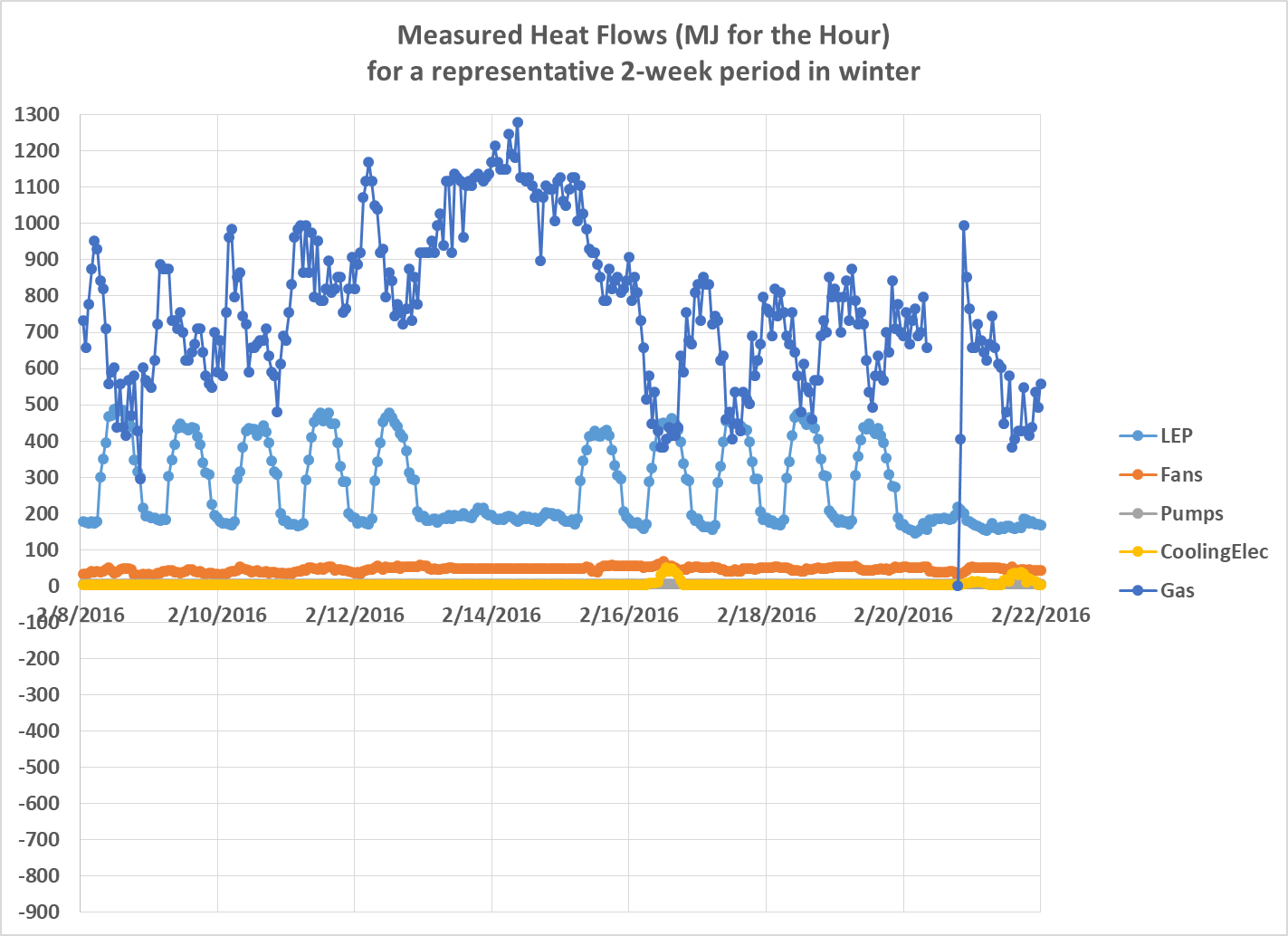}
\label{fig8}
\end{figure}
\begin{figure}[htbp]
\caption{Measured data for the actual building. Only a representative summer period is shown}
\centering
\includegraphics[width=0.8\textwidth]{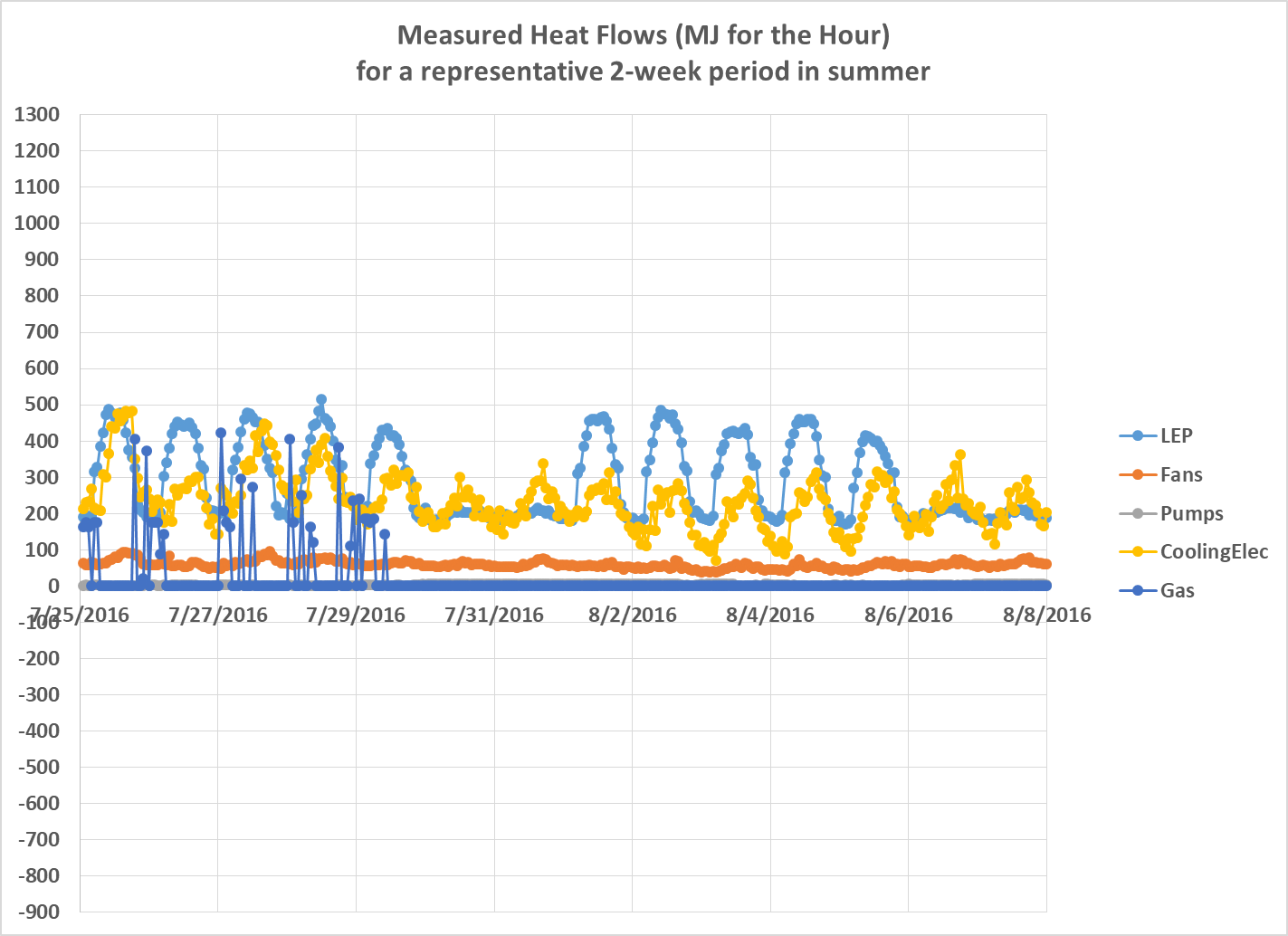}
\label{fig9}
\end{figure}

\begin{figure}[htbp]
\caption{Computed Heat Flows (shown for 2 weeks in winter)}
\centering
\includegraphics[width=0.8\textwidth]{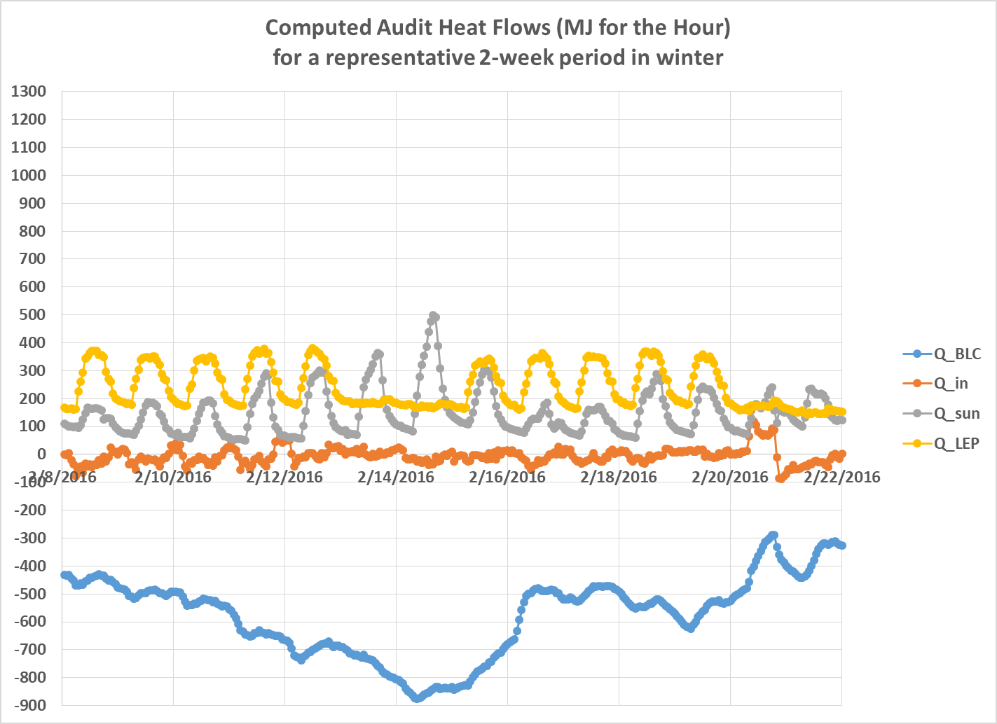}
\label{fig10}
\end{figure}

\begin{figure}[htbp]
\caption{Some computed heat flows (shown for two summer weeks)}
\centering
\includegraphics[width=0.8\textwidth]{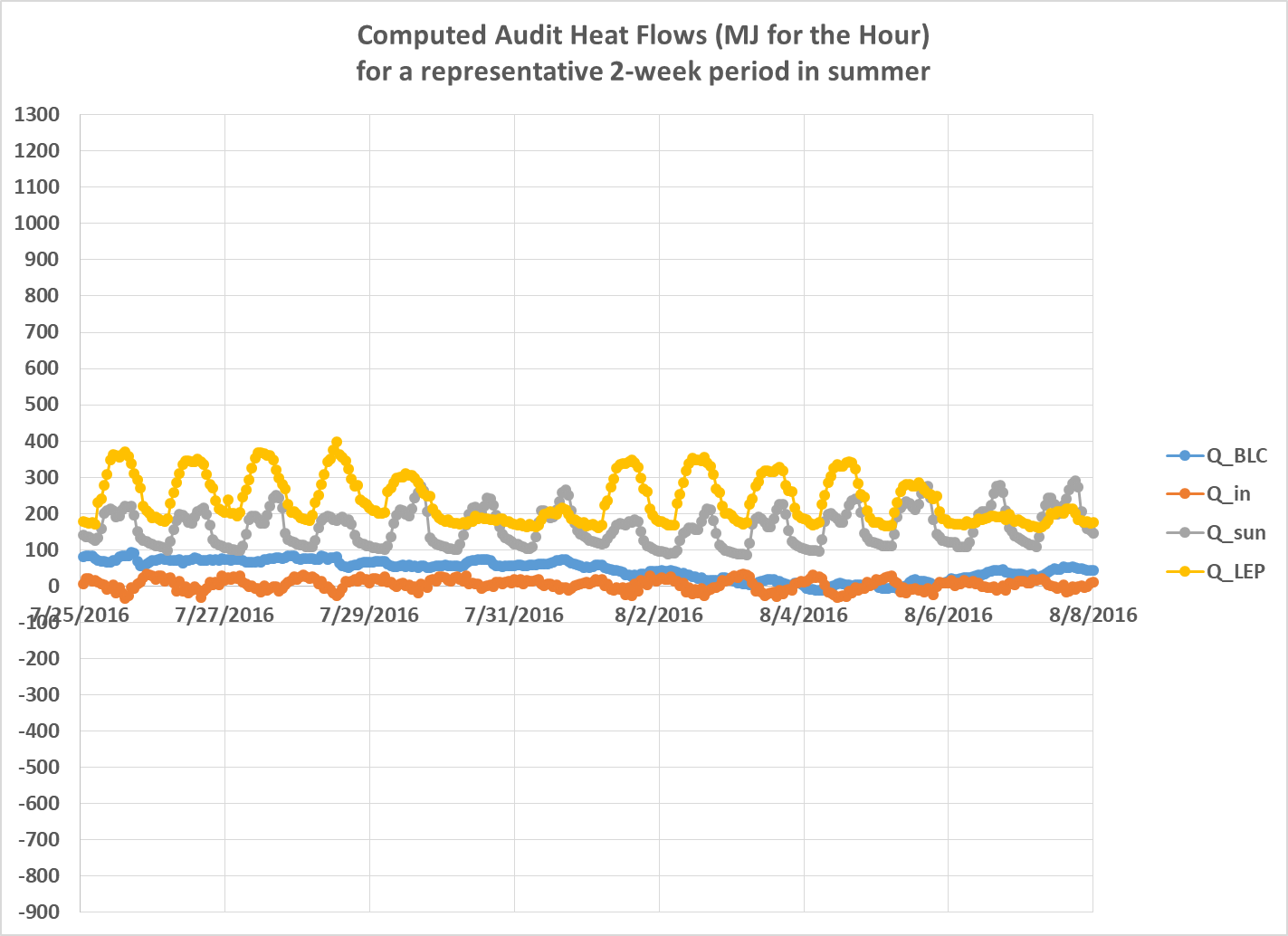}
\label{fig11}
\end{figure}

The building has 7-zones. The measured data is shown, for representative winter and summer periods, in Figure \ref{fig8} and Figure \ref{fig9}. For ease of comparing magnitudes of various heat flows, which is a crucial step, a common scale is used in the graphs. Although some data channels are available for each zone or for each air handler, for the purposes of this graph the sum is presented. LEP represents the energy due to lights, equipment and people. Delayed release of this energy, loss of some energy incident radiatively on exterior walls etc. are considered later during analysis. Similarly, measured gas input to the boiler and electrical energy input to the HVAC systems are shown. Efficiency considerations are also done later during the analysis.
From the 7-zone input file, the four audit heat flows $Q_{BLC}(t)$, $Q_{in}(t)$, $Q_{sun}(t)$, $Q_{LEP}(t)$ are computed for each zone as explained before. In these computations, each zone is scheduled to maintain its measured temperature at each hour. The audit heat flows are shown in Figure \ref{fig10} Computed Heat Flows (shown for 2 weeks in winter)and Figure \ref{fig11} for the summer and winter periods respectively. The varying magnitudes of the individual heat flows both diurnally as well as seasonally are noteworthy.
Additionally, ventilation is estimated from each of the fan flows and associated return, outdoor and mixed air temperatures. Infiltration is modeled following \cite{Infil}

The next step of parameter estimation requires great care. 
\begin{itemize}
    \item 
One can attempt to perform energy balance for each zone and estimate parameters at individual zone level. Given the uncertainties in delivered heating/cooling for each zone and interzonal heat transfer and airflow, this was deemed too problematic. We concluded that one should first perform the energy balance by summing over zones, even though certain flows are calculated at zone level.
    \item
If we have a known (i.e. measured) heat flow (not just gas input or electrical input to a HVAC system), whose magnitude was large enough for at least for some portion of the data period, the parameter estimation would be more robust.  Otherwise, only combinations of parameters can be estimated. A simple example serves to illustrate this: In the Princeton Scorekeeping method (PRISM) analysis [18] for gas heated homes, only the building load coefficient divided by the furnace efficiency can be deduced by fitting the utility bills with variable base degree day data. For our building, delivered heating/cooling energy use was not available. Inspection of the graphs (at least during the plotted period) shows that no suitable period with a large enough known heat flow, even if we were to treat $Q_{LEP}$ as known. (Note: $Q_{LEP}$ is the heat released to the air node after accounting for delayed release, loss through exterior walls etc. It is not equal to $LEP$ input; even the total $Q_{LEP}$ is less than $LEP$ by an amount that depends on the details of radiative couplings as can be seen from the data in the graphs. So $Q_{LEP}$, the actual heat input to the air node, is close to the known electricity consumption by lights and equipment plus heat attributed to people; that is as far as this can be estimated. One can, as an approximation, treat $Q_{LEP}$ as a known heat flow. This is tantamount to accepting that the simulation correctly handles delays and losses of $LEP$. We can then filter on periods when it has large numerical values.)
    \item
As proof of concept, we filtered for periods when $Q_{BLC}$ is high and $Q_{LEP}$,$Q_{sun}$,$Q_{in}$ are low, and found 6 contiguous hours from midnight to 6 am on February 15 to be suitable. The heat flows for that period are shown in Figure \ref{fig12}. The energy balance equation for each hour in this period is 
\begin{multline}
    p_{BLC}Q_{BLC}(t)+p_{BoilerEff}Q_{gas}(t)+\\
    Q_{LEP}(t)+Q_{sun}(t)+Q_{in}(t)+Q_{ventilation}(t)+Q_{infiltration}(t)\approx0
\end{multline}
\begin{figure}[htbp]
\caption{Heat flows, gas, and electricity use for a 6-hour period}
\centering
\includegraphics[width=0.8\textwidth]{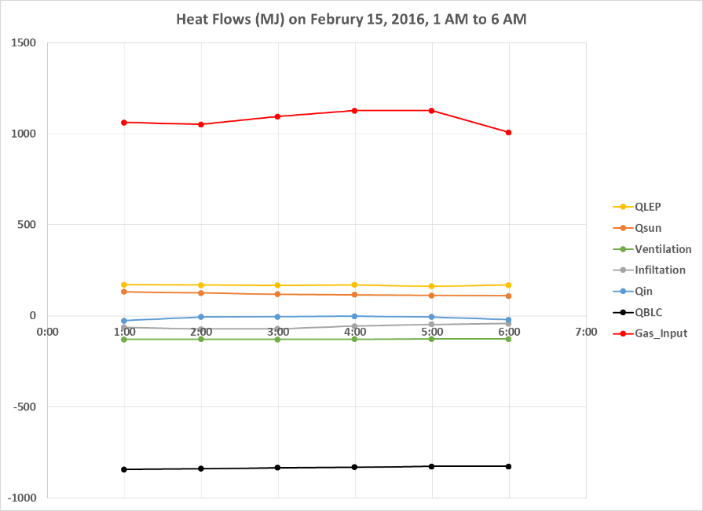}
\label{fig12}
\end{figure}
Because of the relatively low magnitude of $Q_{LEP}$, $Q_{sun}$, $Q_{in}$ in the early morning hours, they are secondary terms and we are justified in not associating parameters with them. We can thereby determine a relationship between $p_{BLC}$, $p_{BoilerEff}$. Knowing one parameter would allow the other to be determined. As we vary $p_{BLC}$, the corresponding value of $p_{BoilerEff}$ can be determined by least squares fit. The resulting relationship is shown in Figure \ref{fig13}
\begin{figure}[htbp]
\caption{Boiler Efficiency as a function of $p_{BLC}$}
\centering
\includegraphics[width=0.8\textwidth]{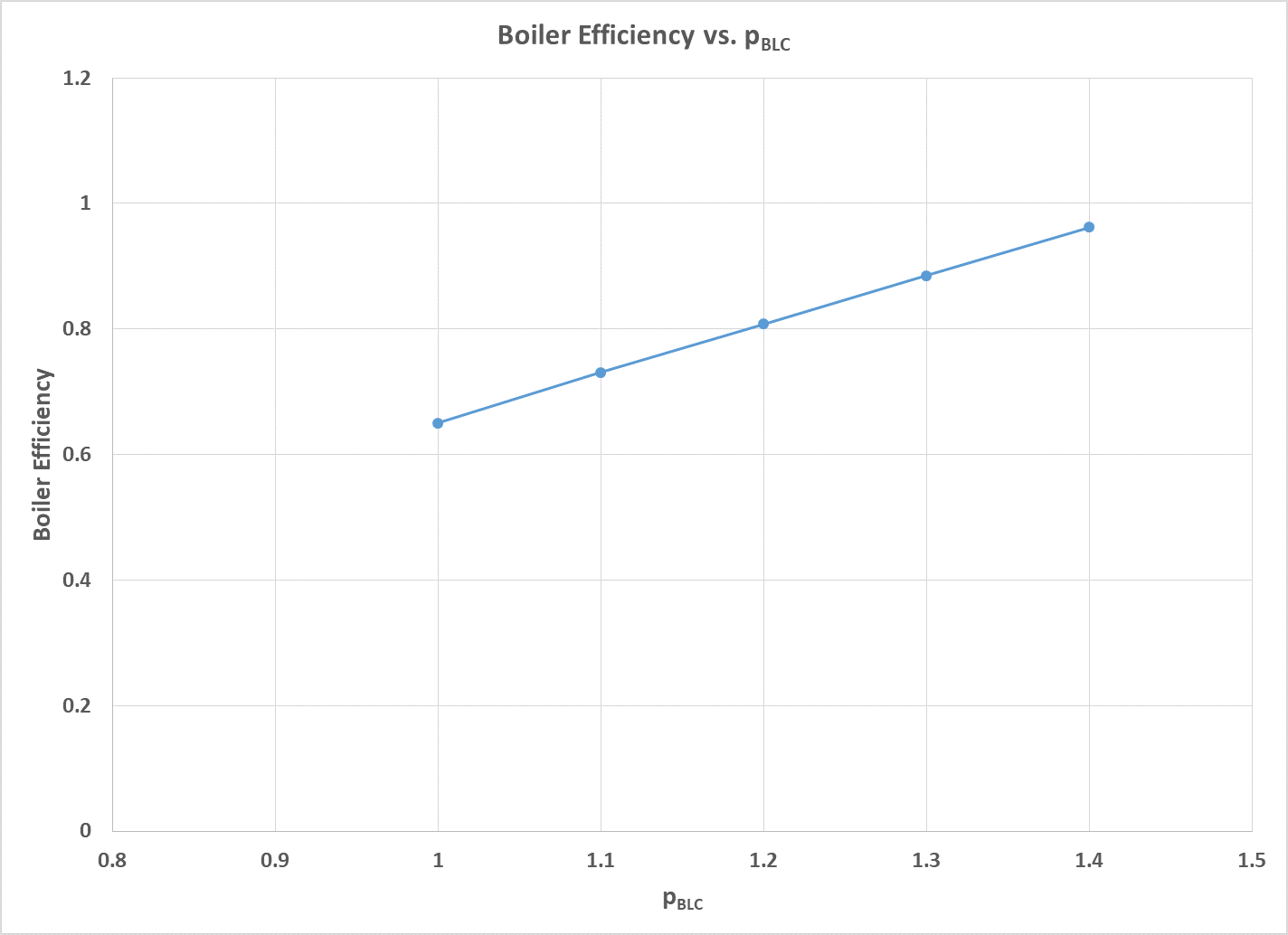}
\label{fig13}
\end{figure}
\end{itemize}

To summarize,  the dominant heat flow is $p_{BLC}Q_{BLC}$ for the time period selected. The terms $Q_{LEP}$, $Q_{sun}$, $Q_{in}$ are small and contributed  to improve the accuracy of $p_{BLC}$ estimate. These heat flows constitute a net heat loss that is compensated by the heat introduced by the boiler with the measured gas input. For any given boiler efficiency (with default part-load and off-rated curves) we can determine the heat introduced by the boiler. This relationship is shown in Figure \ref{fig13}. If we assume the boiler efficiency to be in the 80\% to 90\% range, $p_{BLC}$, will be in the 1.2 to 1.3 range, implying that the envelope is 20\% to 30\% more lossy than implied by the EnergyPlus input file.

The subsequent steps, involving introduction of the usual additional transfer function parameters and training a neural net for the residuals, required data that was not available to us.

\section{Summary and Future Work}

The Enhanced parameter Estimation (EPE) methodology described in this paper is significantly different from conventional approaches to calibration. It allows one to systematically and objectively reconcile building energy simulations with actual performance data.  In summary, EPE consists of the following steps:
\begin{itemize}
\item	Various sources of zone level heat flows are defined and computed using a set of specially formulated EnergyPlus simulation runs
\item	Least Squares fit to the model of the energy balance is accomplished by introducing and estimating parameters with physical significance. This aspect has been demonstrated in both synthetic and real buildings 
\item	A two-stage calibration method is developed: (i) for analyzing the building shell, the HVAC system is replaced by an ideal system; and (ii) for analyzing the HVAC system, the building shell is replaced by a box with only process loads.
\item	Improved model prediction accuracy is achieved by training a neural net to the time series of residuals remaining after the least squares parameter estimation.
\item	Inputs to the neural net are computed heat flows that automatically incorporate the history of temperatures, solar radiation, internal gains, etc., thus simplifying a time series problem into one in which each hour is independent as far as the neural net is concerned.

\end{itemize}

Future work is planned to address (a) incorporation of the parameters into EnergyPlus simulation with thermostatic constraints, (b) refinements to multizone buildings, (c) automatic determination of data windows in which different parameters are best elicited through classification methods, (d) advanced HVAC parameter estimation (e) refining the methodology with data from additional actual buildings.

\section{Acknowledgments}
The authors gratefully acknowledge an SBIR grant from the Department of Energy, with Dr. Amir Roth as Program Manager. Dr James Freihaut and Mr. Scott Wagner of Pennsylvania State University kindly provided building data and a simulation model for the actual building used to illustrate/evaluate the EPE methodology.

\printbibliography
\end{document}